\documentclass[useAMS,usenatbib]{mn2e} 
\usepackage{graphicx} 
\usepackage[usenames]{color} 
\newcommand{\changea}{\textcolor{Blue} }

\title[Physical parameters of {\it Kepler} target]{Atmospheric 
parameters and pulsational properties for a sample of $\delta$\,Sct,
$\gamma$\,Dor, and hybrid {\it Kepler} targets\thanks{This work 
is based on spectra taken at the Loiano (INAF\,-\,OA Bologna) and Serra La
Nave (INAF\,-\,OA Catania) Observatories.}} 
\author[G. Catanzaro et al.]{G. Catanzaro$^{1}$\thanks{E-mail: gca@oact.inaf.it}, 
V. Ripepi$^{2}$, 
S. Bernabei$^{3}$, 
M. Marconi$^{2}$, 
L. Balona$^{4}$, 
\and D. W. Kurtz$^{5}$, 
B. Smalley$^{6}$, 
W. J. Borucki$^{7}$,
H. Bruntt$^{8,9}$, 
J. Christensen-Dalsgaard$^{9}$,
\and A. Grigahc{\`e}ne$^{10}$,
H. Kjeldsen$^{9}$,
D. G. Koch$^{7}$,
M. J. P. F. G. Monteiro$^{10}$,
\and J.C. Su\'arez$^{11}$,
R. Szab\'o$^{12}$,
K. Uytterhoeven$^{13}$
\\  
$^{1}$INAF-Osservatorio Astrofisico di Catania, Via S.Sofia 78, I-95123, Catania, Italy\\ 
$^{2}$INAF-Osservatorio Astronomico di Capodimonte, Via Moiariello 16, I-80131, Napoli, Italy\\ 
$^{3}$INAF-Osservatorio Astronomico di Bologna, Via Ranzani 1, I-40127, Bologna-Italy\\ 
$^{4}$South African Astronomical Observatory, P.O. Box 9, Observatory 7935, Cape Town, South Africa\\ 
$^{5}$Jeremiah Horrocks Institute of Astrophysics, University of Central Lancashire, Preston PR1\,2HE, UK\\ 
$^{6}$Astrophysics Group, Keele University, Keele, Staffordshire ST5 5BG, UK\\ 
$^{7}$NASA Ames Research Center, MS 244-30, Moffett Field, CA 94035, USA\\
$^{8}$Observatoire de Paris, LESIA, 5 place Jules Janssen, 92195 Meudon Cedex, France\\ 
$^{9}$Department of Physics and Astronomy, Building 1520, Aarhus University, 8000 Aarhus C, Denmark\\
$^{10}$Centro de Astrof\'{\i}sica and Faculdade de Ci\^encias, Universidade do Porto, Rua das Estrelas, 
      410-762 Porto, Portugal\\ 
$^{11}$Instituto de Astrof\'{\i}sica de Andaluc\'{\i}a (CSIC). Rotonda de la Astronom\'{\i}a S/N. Granada, Spain\\
$^{12}$Konkoly Observatory of the Hungarian Academy of Sciences, PO Box 67, 1525 Budapest, Hungary\\ 
$^{13}$Laboratoire AIM, CEA/DSM-CNRS-Universit\'e Paris Diderot; CEA, IRFU, SAp, centre de Saclay, 91191, Gif-sur-Yvette, France\\
} 
 
\date{Accepted 2010 September 21. Received 2010 September 21; in original form 2010 June 4} 
 
\pagerange{\pageref{firstpage}--\pageref{lastpage}} \pubyear{2002} 
 
\def\LaTeX{L\kern-.36em\raise.3ex\hbox{a}\kern-.15em 
 T\kern-.1667em\lower.7ex\hbox{E}\kern-.125emX} 
 
\begin{document} 
 
\label{firstpage} 
 
\maketitle 
 
\begin{abstract} 
We report spectroscopic observations for 19 $\delta$\,Sct candidates observed
by the {\it Kepler} satellite both in long and short cadence mode. For all these 
stars, by using spectral synthesis, we derive the effective temperature, 
the surface gravity and the projected rotational velocity. An equivalent spectral 
type classification has been also performed for all stars in the sample. These 
determinations are fundamental for modelling the frequency spectra that will be 
extracted from the {\it Kepler} data for asteroseismic inference. 
For all the 19 stars, we present also periodograms obtained from {\it Kepler} 
data. We find that all stars show peaks in both low- ($\gamma$\,Dor; 
g mode) and high-frequency ($\delta$\,Sct; p mode) regions. Using the amplitudes and
considering 5\,c/d as a boundary frequency, we classified 3 stars as pure $\gamma$\,Dor, 
4 as $\gamma$\,Dor\,-\,$\delta$\ hybrid, Sct, 5 as
$\delta$\,Sct\,-\,$\gamma$\,Dor hybrid,
and 6 as pure $\delta$\,Sct. The only exception is the star KIC\,05296877 
which we suggest could be a binary.

\end{abstract} 
 

 \begin{table*} 
\caption{List of studied {\it Kepler} $\delta$\,Sct candidates. Columns (1) and 
(2) show the {\it Kepler} and other identifications; Column (3) reports the origin 
of the spectrum: L\,=\,Loiano, OACT\,=\,INAF\,-\,OACT; Columns (4), (5), and (6) 
report the $V$ magnitudes, the colour excess, and the spectral types taken from the 
literature.} 
\begin{center} 
\begin{tabular}{cllrcl}  
\hline 
\hline 
\noalign{\medskip} 
    KIC ID &  ~~~Other ID  & Observ.&  $V$~ &      $E(B-V)$        & S.T.   \\ 
\noalign{\medskip}  &       &        &  mag  &      mag             &        \\ 
(1)        & ~~~~~(2)      & ~~(3)  & (4)   &         (5)          & (6)    \\ 
\noalign{\medskip} 
\hline 
\noalign{\smallskip} 
03219256  & HD\,178306     & L,OACT &  8.31 &  0.02$\pm$0.01$^{a}$ & A3$^{1}$      \\  
03429637  & HD\,178875     & OACT   &  7.72 &  0.04$\pm$0.02$^{b}$ & A9,Am$^{2}$   \\ 
03437940  & HD\,181569     & L,OACT &  8.45 &  0.06$\pm$0.01$^{a}$ & F0$^{1}$      \\  
04570326  & HAT 199-01905  & L      &  9.77 &  0.00$\pm$0.02$^{c}$ & A2$^{3}$      \\ 
05296877  & HAT 199-27597  & L      & 12.50 &  0.00$\pm$0.02$^{c}$ &               \\  
05724440  & HD\,187234     & L,OACT &  7.90 &  0.06$\pm$0.01$^{a}$ & A5$^{3}$      \\  
05965837  & HAT 199-00623  & L      &  9.23 &  0.01$\pm$0.02$^{c}$ & F2$^{1}$      \\  
05988140  & HD\,188774     & L      &  8.83 &  0.07$\pm$0.01$^{a}$ & F0$^{1}$      \\  
07119530  & HD\,183787     & L      &  8.45 &  0.02$\pm$0.01$^{a}$ & A3$^{1}$      \\   
07798339  & HD\,173109     & L,OACT &  7.87 &  0.00$\pm$0.01$^{a}$ & F0$^{4}$      \\  
08197788  & NGC6866-V1     & L      & 12.98 &  0.12$\pm$0.02$^{d}$ &               \\ 
08264404  & NGC6866-V3     & L      & 12.22 &  0.12$\pm$0.02$^{d}$ &               \\ 
08264698  & NGC6866-V2     & L      & 12.38 &  0.12$\pm$0.02$^{d}$ &               \\ 
08583770  & HD\,189177     & L      & 10.16 &  0.17$\pm$0.03$^{c}$ & B9$^{5}$      \\  
09655114  & NGC6811-RH35   & L      & 12.09 &  0.12$\pm$0.02$^{e}$ & A4$^{6}$      \\ 
09775454  & HD\,185115     & L,OACT &  8.19 &  0.01$\pm$0.01$^{a}$ & F1IV$^{7}$    \\  
11402951  & HD\,183489     & L,OACT &  8.14 &  0.03$\pm$0.01$^{b}$ & A9,Am$^{2}$   \\  
11445913  & HD\,178327     & L,OACT &  8.46 &  0.03$\pm$0.01$^{b}$ & A9,Am$^{2}$   \\  
11973705  & HD\,234999     & L      &  9.10 &  0.00$\pm$0.02$^{c}$ & B9$^{1}$      \\ 
\noalign{\smallskip}  
\hline 
\end{tabular}  
\end{center} 
\begin{flushleft} 
{\small a) from $uvby\beta$ photometry; b) from ($b-y$) photometry; c)
  from ($B-V$) photometry; d) \citet{dutra}; e) \citet{Glushkova}.}  \\
{\small 1) SIMBAD; 2) \citet{abt}; 3) \citet{feh}; 4) \citet{nordstrom}; 5) \citet{cauteau}; 
6) \citet{lindoff}; 7) \citet{duflot}} 
\end{flushleft} 
\label{tab1} 
\end{table*}

\section{Introduction} 
Stellar pulsations offer a unique opportunity to constrain the intrinsic 
parameters of stars and, by using asteroseismology, to unveil their inner 
structure. In particular, the classical $\delta$\,Sct variables are late A-type 
and early F-type stars that populate the instability strip between the 
zero-age main sequence and terminal-age main sequence, with $3.0 \le {\rm M}_V \le 
0.5$. They pulsate in mode of low radial order with periods ranging from about 20\,min to 8\,h
\citep[see][for a review]{breger00}. 
$\delta$\,Sct stars pulsate in both radial and non-radial p\,modes and g\,modes, driven by the 
$\kappa$-mechanism, in particular in the He\,\textsc{ii} ionization zone. The $\gamma$\,Dor 
variables, with periods between about 0.3 and 3\,d, are mostly located near the 
cool edge of the $\delta$\,Sct instability strip \citep{kaye99}. Their pulsations are driven by 
convective blocking at the base of their envelope convection zone
\citep{guzik00, dupret04, grig04}. The distinction between the two classes is clearer if we consider the 
value of the pulsation constant, Q \citep{handler}. However, the location of the $\gamma$\,Dor 
stars in the Hertzsprung-Russell (HR) diagram suggests some 
relationship with the $\delta$\,Sct variables. Indeed, stars exist which show simultaneously 
both $\delta$\,Sct and $\gamma$\,Dor pulsations \citep{henry05,
  king06, rowe06, uytte08, handler09}. These hybrid objects are in principle of great interest, 
because they offer additional constraints on stellar
structure. Indeed, the $\gamma$\,Dor stars pulsate in g\,modes which have high amplitudes deep in the 
star and allow us to probe the stellar core, while the p\,modes, efficient in 
$\delta$ Sct stars,  have high amplitudes in the outer regions of the star and 
probe the stellar envelope. Moreover, since $\gamma$\,Dor stars pulsate in g\,modes of 
high radial order, the asymptotic approximation predicts regular patterns in the periods 
which, in principle, can be used to determine the spherical harmonic degree. Hybrids thus 
provide a unique opportunity for asteroseismology which is not available in pure 
$\delta$\,Sct stars \citep{handler}. Recently, a large separation-like feature
has been discovered in the $\delta$\,Sct star HD\,174936 by \citet{garcia09}
using CoRot data. In that work, such a regularity was used to constrain the 
modelling of this star. It is extremely interesting to search for such
regularities in hybrid stars, for which the presence of other frequency 
regimes definitely may help the overall comprehension of the pulsational 
behaviour of these objects.
 
The {\it Kepler} satellite\footnote{ http://kepler.nasa.gov/} was launched on 
2009~March~6 and will continuously monitor the brightness of over 100\,000 stars 
for at least 3.5\,yr in a 105 square degree fixed field of view near the 
plane of the Milky Way between Deneb and Vega. The main aim of the mission is to 
detect extrasolar planets, particularly Earth-sized planets in the 
habitable zone of their stars, where water may be liquid, by the transit method \citep{borucki97}. 
To accomplish this goal, {\it Kepler} is capable of measuring the stellar brightnesses to 
$\mu$mag precision \citep{gilliand10} which, together with the long duration of the 
observations, make the data ideal for asteroseismology. Most of the observations are 
long-cadence (29.4-min) exposures, though a small allocation is available for short-cadence (1-min) 
exposures. The long-cadence as well as some short-cadence data released to the 
{\it Kepler Asteroseismic Science Consortium} (KASC) have been surveyed for 
$\delta$\,Sct and/or $\gamma$\,Dor stars. The long-cadence data are not always suitable 
for a detailed study of $\delta$\,Sct oscillations because many of these stars have 
frequencies higher than the Nyquist frequency (24.5\,c/d) for 29.4-min sampling. 
These data are, however, suitable for the  detection of $\delta$\,Sct -- $\gamma$\,Dor hybrids. 
 
The early KASC data releases led to the discovery of the nineteen
candidate $\delta$\,Sct stars listed in Table~\ref{tab1}; 
many more have been found in subsequent data releases, and ground-based 
studies of these stars are now in progress. Interestingly, many objects show 
periodograms with frequencies both in the p\,mode $\delta$\,Sct and g\,mode $\gamma$\,Dor 
domains, i.e., they are candidate hybrid pulsators \citep{griga10}. Dedicated short-cadence 
{\it Kepler} data for the most promising hybrid candidates will be exploited for seismic studies 
of these stars. To this end it is extremely important to constrain the fundamental parameters 
of the stars  (effective temperature $T_{\rm eff}$, surface gravity $\log g$, projected 
rotational velocity $v \sin i$, luminosity $\log L/{\rm L}_\odot$) in order to limit 
the range of models. Measurement of  $v \sin i$ is essential to constrain the rotational 
velocity of the models. Stellar fundamental parameters can be obtained by using photometry, 
e.g., in the Str\"omgren system, or by means of mid- or high-resolution spectroscopic 
observations. Very few of the 19 {\it Kepler} $\delta$\,Sct stars have previously  been observed 
spectroscopically and no reliable estimates of the stellar parameters can be derived from the existing 
data. For this reason, we undertook a systematic spectroscopic study of these {\it Kepler} targets and 
report our results here. This work fits in the  ground-based observational efforts of KASC with the  aim to characterize all {\it Kepler} pulsators \citep{uytte10a,uytte10b}.

\begin{table*} 
\caption{Comparison of fundamental parameters obtained from spectroscopy and 
photometry, columns (2)-(6). In column (2) we report the Equivalent Spectral Type, i.e.,
the spectral type assigned to the stars by comparing the spectroscopic $T_{\rm eff}$
with the table in \citet{schmidt}} 
\begin{center} 
\begin{tabular}{llrcccccccc} 
\hline 
\hline 
\noalign{\medskip} 
KIC ID  & E. S.T. & $v \sin i$~~   & $T^{\rm Spec}_{\rm eff}$  & $T^{(V-K)}_{\rm eff}$ & $T^{uvby\beta}_{\rm eff}$ & $T^{\rm IRFM}_{\rm eff}$ & $T^{\rm KIC}_{\rm eff}$ & $\log g^{\rm Spec}$  & $\log g^{uvby\beta}$ &$\log g^{KIC}$\\ 
\noalign{\medskip} 
      &         & km s$^{-1}$    & K                         &     K                 &      K                &           K                  &                         &  cm/s$^{2}$          & cm/s$^{2}$    & cm/s$^{2}$        \\ 
(1)   &  (2)    & (3)            & (4)                       &    (5)                &    (6)                &       (7)                    &  (8)                    &     (9)              &  (10)   &  (11)          \\ 
\noalign{\medskip} 
\hline 
\noalign{\smallskip}                                                 
03219256   & A8V    &  90$\pm$5  & 7500$\pm$150 & 7450$\pm$150 &  7650$\pm$250  & 7450$\pm$150 &  7300$\pm$250  &  3.6$\pm$0.1 & 4.1$\pm$0.3   &  3.6$\pm$0.3 \\
03429637   & F0III    &  50$\pm$5  & 7100$\pm$150 & 7650$\pm$200 &                & 7400$\pm$150 &  6950$\pm$250  &  3.0$\pm$0.2 &               &  3.4$\pm$0.3 \\
03437940   & A7.5V  & 120$\pm$5  & 7700$\pm$120 & 7800$\pm$150 &  7850$\pm$250  & 7900$\pm$150 &  7400$\pm$250  &  4.1$\pm$0.2 & 3.7$\pm$0.3   &  3.9$\pm$0.3 \\
04570326   & F1V    &  80$\pm$20 & 7000$\pm$150 & 6600$\pm$200 &                & 6900$\pm$150 &                &  4.0$\pm$0.3 &               &               \\
05296877   & F4.5IV & 200$\pm$20 & 6500$\pm$200 & 6000$\pm$400 &                & 6500$\pm$150 &  6650$\pm$250  &  3.8$\pm$0.3 &               &  4.1$\pm$0.3 \\
05724440   & A9IV   & 220$\pm$5  & 7350$\pm$120 & 7700$\pm$150 &  8150$\pm$250  & 7850$\pm$150 &  7300$\pm$250  &  3.6$\pm$0.3 & 4.2$\pm$0.3   &  3.6$\pm$0.3 \\
05965837   & F1V    &  20$\pm$10 & 6975$\pm$200 & 6700$\pm$200 &                & 6850$\pm$150 &  6500$\pm$250  &  4.0$\pm$0.4 &               &  4.1$\pm$0.3 \\
05988140   & A9IV   &  70$\pm$20 & 7400$\pm$150 & 7750$\pm$150 &  7900$\pm$250  & 7900$\pm$150 &  7450$\pm$250  &  3.7$\pm$0.3 & 3.7$\pm$0.3   &  3.5$\pm$0.3 \\
07119530   & A8IV   & 200$\pm$20 & 7500$\pm$200 & 7750$\pm$150 &  7800$\pm$250  & 7850$\pm$150 &  7800$\pm$250  &  3.6$\pm$0.3 & 3.3$\pm$0.3   &  3.5$\pm$0.3 \\
07798339   & F3IV   &  15$\pm$5  & 6700$\pm$200 & 6900$\pm$150 &  6800$\pm$250  & 6900$\pm$150 &  6700$\pm$250  &  3.7$\pm$0.3 & 3.7$\pm$0.3   &  3.4$\pm$0.3 \\
08197788   & A8V    & 230$\pm$20 & 7500$\pm$150 & 7700$\pm$200 &                & 7950$\pm$150 &  8000$\pm$250  &  4.0$\pm$0.3 &               &  4.0$\pm$0.3 \\
08264404   & A8IV   & 250$\pm$20 & 7500$\pm$200 & 7500$\pm$200 &                & 7950$\pm$150 &  7900$\pm$250  &  3.7$\pm$0.3 &               &  3.7$\pm$0.3 \\
08264698   & A8V    & 210$\pm$20 & 7500$\pm$200 & 7650$\pm$200 &                & 8000$\pm$150 &  8000$\pm$250  &  3.9$\pm$0.2 &               &  3.8$\pm$0.3 \\
08583770   & A2III  & 130$\pm$10 & 9000$\pm$200 & 8200$\pm$300 &                & 8750$\pm$200 &  7650$\pm$250  &  3.0$\pm$0.2 &               &  3.5$\pm$0.3 \\
09655114   & A9V    & 150$\pm$20 & 7400$\pm$200 & 7850$\pm$250 &                & 7700$\pm$150 &  7750$\pm$250  &  3.9$\pm$0.3 &               &  3.8$\pm$0.3 \\
09775454   & F1V    &  70$\pm$5  & 7050$\pm$150 & 7150$\pm$150 &  7300$\pm$250  & 7050$\pm$150 &                &  4.0$\pm$0.3 & 4.0$\pm$0.3   &              \\
11402951   & F0IV   & 100$\pm$5  & 7150$\pm$120 & 7450$\pm$150 &                & 7450$\pm$150 &                &  3.5$\pm$0.1 &               &              \\
11445913   & F0IV   &  50$\pm$5  & 7200$\pm$120 & 7150$\pm$150 &                & 7150$\pm$150 &  6950$\pm$250  &  3.5$\pm$0.2 &               &  3.9$\pm$0.3 \\
11973705   & A9.5V  & 120$\pm$20 & 7300$\pm$300 & 7350$\pm$200 &                & 7400$\pm$150 &  7400$\pm$250  &  4.2$\pm$0.3 &               &  4.0$\pm$0.3 \\
\noalign{\smallskip} 
\hline 
\end{tabular} 
\end{center} 
\label{tab2} 
\end{table*}

\section{Observations and data reduction} 
\label{obser} 
The spectra used in our analysis were acquired with two different instruments: 
 
\begin{enumerate} 
\item {\it Loiano Observatory:} We used the Bologna Faint Object Spectrograph \& 
Camera (BFOSC) instrument attached to the 1.5-m Loiano 
telescope\footnote{http://www.bo.astro.it/loiano/index.htm}. We adopted 
the echelle configuration with Grism \#9 and \#10 (as cross dispersers). The 
typical resolution was $R\sim 5000$. Spectra were recorded on a back-illuminated 
(EEV) CCD with $1300 \times 1300$ pixels of 20\,$\mu$m size, typical readout noise 
of 1.73 e$^{-}$ and gain of 2.1\,e$^-$/ADU. Observations were carried out during 
the nights of 2009 September $8 - 11$. The exposure times ranged from 1200 to 
3600\,s for the brightest and faintest targets respectively. 
 
\item {\it INAF\,-\,OACT:} the 91-cm telescope of the Istituto Nazionale 
di Astrofisica -- Osservatorio Astrofisico di Catania (INAF--OACT) was used to 
carry out spectroscopy of eight of the targets. The telescope is fibre linked to a 
REOSC\footnote{REOSC is the Optical Department of the SAGEM Group.} 
echelle spectrograph, giving R\,$\sim$\,20000 spectra in the range $4300 -
6800$\,\AA. The resolving power was checked using the Th-Ar calibration lamp. 
Spectra were recorded on a thinned, back-illuminated (SITE) CCD with $1024 \times 
1024$ pixels of 24\,$\mu$m size. The typical readout noise was 6.5 e$^{-}$ at a 
gain of 2.5\,e$^-$/ADU. Observations were carried out during five nights: 2009 
September 28, 29, 30 and November 19 and 28. Exposure times were fixed for all the 
stars at 1\,h. 
\end{enumerate} 
 
The reduction of spectra, which included the subtraction of the bias frame, 
trimming, correcting for the flat-field and the scattered light, the extraction of 
the orders, and the wavelength calibration, was done by using the NOAO/IRAF 
package\footnote{IRAF is distributed by the National Optical Astronomy 
Observatory, which is operated by the Association of Universities for Research in 
Astronomy, Inc.}. The amount of scattered light correction was about 10\,ADU. The 
S/N ratio of the spectra was at least $\sim$\,130 and 80 for Loiano and OACT 
observatories, respectively.

\section{Physical parameters} 
\label{parameter}

\subsection{Parameters from photometry: $T_{\rm eff}$ and  $\log g$} 
 
Complete Str\"omgren-Crawford $uvby\beta$ photometry is available for seven objects in our 
sample (stars with an ``$a$'' in column 5 of Table\,\ref{tab1}) while $uvby$ data 
are present for three additional objects (stars with a ``$b$'' in column 5 of 
Table\,\ref{tab1}). The source of both the Str\"omgren and Str\"omgren-Crawford data 
is \citet{hauck}. For the other six objects (identified with a ``$c$'' in column 5 of Table\,\ref{tab1})
plus the three stars in NGC\,6866, only Johnson photometry is available, mainly in 
$BV$ filters. For these stars we used the values reported by {\it SIMBAD}. In the 
near-infrared, $JHK$ photometry of good quality is present in the 2MASS catalogue 
\citep{2mass} for all the targets. 
 
For the seven stars with $uvby\beta$ photometry, $E(b-y)$ can be estimated by using the 
calibration by \citet{moon}, using the {\it IDL} code \textsc{uvbybeta}. 
The result is reported in Table~\ref{tab1}, where we have used the transformation 
$E(B-V)=1.4\,E(b-y)$ \citep{cardelli89}. For the three stars without $\beta$ indices 
we used the equivalent spectral type derived in Section~\ref{parameters_from_spectroscopy} 
to derive the intrinsic $(b-y)$ from the relation between $(b-y)_0$ and spectral type \citep{born06}. 
Similarly, for five stars with $(B-V)$ colours, we adopted the intrinsic 
$(B-V)_0$ colours as a function of spectral type \citep{schmidt}. We assigned a 
larger error to these values than those based on $uvby\beta$ photometry. 
The remaining four variables are cluster members, three belong to NGC 6866 and one 
to NGC 6811. The reddenings of these two clusters were adopted from \citet{dutra}  
for NGC 6866 and \citet{Glushkova} and \citet{luo} for NGC 6811.

Values of $T_{\rm eff}$ and $\log g$ were estimated from $c_0$ and $\beta$ using 
the data grid by \citet{moondwo}. 
Typical photometric errors (0.015 and 0.03\,mag in $\beta$ and $c_0$, 
respectively) have been assumed. The derived values of $T_{\rm eff}$ and $\log g$ 
are reported in Table\,\ref{tab2} (columns 6 and 10, respectively). In calculating the 
uncertainties, we conservatively assumed an error of 250\,K and 0.3\,dex for 
$T_{\rm eff}$ and $\log g$, respectively. The values of $T_{\rm eff}$ and $\log g$
derived from spectroscopy (see Section~\ref{parameters_from_spectroscopy} 
below) and $uvby\beta$ are in good agreement, with all the differences less
than 3$\sigma$. Only KIC\,05724440 (HD\,187234) shows a difference in temperature 
close to the edge of this limit. We discuss this object in 
Section~\ref{individuals}.

\begin{figure*} 
\includegraphics[width=12cm,angle=-90]{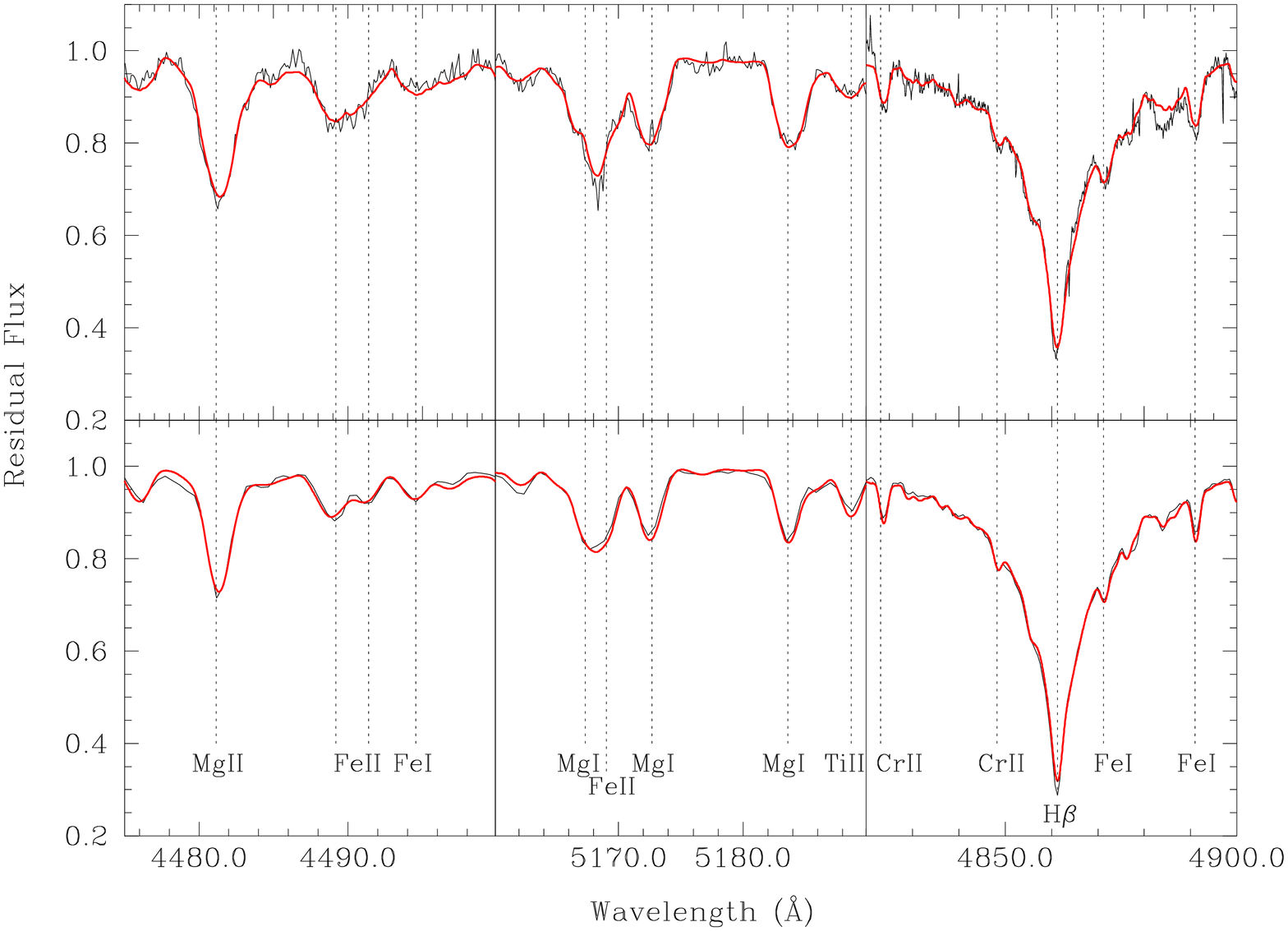} 
\caption{Examples of the fitting procedure for two stars of our sample. The top 
panel shows three spectral regions of KIC\,11402951 (HD\,183489) observed with the OACT equipment, 
while the bottom panel shows the same spectral range, but for KIC\,05988140 (HD\,188774) observed 
with the Loiano spectrograph. Both synthetic spectra have been computed with 
\textsc{synthe} based on LTE \textsc{atlas9} atmospheric models with solar ODF and 
metallicity.} 
\label{spectra} 
\end{figure*}

An additional photometric estimate of $T_{\rm eff}$ can be obtained from the 
calibrations by \citet{masana06}. These are based on the $(V-K)_0$ colour as well 
as on $\log g$ and [$Fe/H$]. The $V$ and $K$ band were taken from the {\it SIMBAD} 
and {\it 2MASS} catalogues, respectively. For $\log g$ we use the value from our 
spectroscopy. For the metallicity, following \citet{bruntt08}, we adopted 
$[Fe/H]=-0.2 \pm 0.2$. This arbitrary value has only a small impact on the results 
because varying $[Fe/H]$ by $2\sigma$ gives an error of only 40\,K in $T_{\rm 
eff}$. To de-redden the observed $(V-K)$ colours we adopted the reddening reported in 
Table\,\ref{tab1}, using the relation $E(V-K) = 3.8\,E(b-y)$ \citep[][]{cardelli89} 
for the stars with Str\"omgren photometry, and $E(V-K) = 2.8\,E(B-V)$ for the 
other stars. The resulting $T_{\rm eff}$ and the relative errors are reported in 
Table\,\ref{tab2} (column 5). In general, there is good agreement between the 
photometric and spectroscopic values. 

Near infrared photometry from 2MASS, complemented by the one in the optical, can be
used to derive an alternative estimate of  $T_{\rm eff}$ by means of
the Infrared Flux Method (IRFM, Blackwell \& Shallis 1977). In particular, 
broad-band photometry (this work, plus TASS4 I-mag, NOMAD R-mag, CMC14 r' mag and 
2MASS photometry) was used to estimate the total observed bolometric flux ($f_{\rm tot}$). 
The photometry was converted to fluxes and the best-fitting \citet{kur93b} model 
flux distribution was found and integrated to determine $f_{\rm tot}$. The 
Infrared Flux Method \citep{black77} was then used with 2MASS fluxes to determine 
the T$_{\rm eff}$ reported in Table~\ref{tab2} (column 7).

Finally, we inspected the {\it Kepler} Input Catalogue (KIC\footnote{accessible via
http://archive.stsci.edu/kepler/kepler\_fov/search.php}, \citet{latham05}) where
additional estimates for $T_{\rm eff}$ and $\log g$ based mainly on 
$u^{'},g^{'},r^{'},i^{'},z^{'}$\footnote{Note, however, that only 25\%
of the KIC stars have  $z^{'}$ photometry, and less than 0.1\% have $u^{'}$ (see
http://nsted.ipac.caltech.edu/data/NStED/kic\_columns.html)}
are present. These value are reported in columns (8) and (11) of
Table~\ref{tab2}. It is worth noticing that the KIC catalogue was
mainly aimed at separating dwarfs from giants, therefore the  
$T_{\rm  eff}$ and $\log g$ are not expected to be very precise. Since
no errors are present in the KIC catalogue, we assumed uncertainties of 250 K
and 0.3 dex in $T_{\rm eff}$ and $\log g$, respectively.

\subsection{Parameters from spectroscopy: $T_{\rm eff}$, $\log g$ and rotational 
velocities}
\label{parameters_from_spectroscopy}
 
We determined $T_{\rm eff}$ and $\log g$ of the stars by minimizing the difference 
between the observed and the synthetic H$\beta$ profiles.  For the goodness-of-fit 
parameter we used $\chi^2$ defined as\\ 
 
$\displaystyle \chi^2 = \frac{1}{N} \sum \bigg(\frac{I_{\rm obs} - I_{\rm 
th}}{\delta I_{\rm obs}}\bigg)^2$ 
 
\bigskip 
 
\noindent 
where $N$ is the total number of points, $I_{\rm obs}$ and $I_{\rm th}$ are the 
intensities of the observed and computed profiles, respectively, and $\delta 
I_{\rm obs}$ is the photon noise. The errors have been estimated from the 
variation in the parameters required to increase $\chi^2$ by one. As starting 
values of $T_{\rm eff}$ and $\log g$, we used $T_{\rm eff}$ and $\log g$ derived 
from the photometry, as described in the previous section. At the 
same time, we determined the projected rotational velocity by matching the Mg{\sc 
ii}~$\lambda$4481\,\AA\ profile with a synthetic profile. The synthetic profiles are computed with 
\textsc{synthe} \citep{kur81} on the basis of \textsc{atlas9} \citep{kur93} LTE 
atmosphere models. All models are calculated using the solar opacity distribution 
function (ODF), solar metallicity and a microturbulence velocity of 
$\xi$\,=\,2\,km\,s$^{-1}$. The atomic parameters for the spectral lines were 
taken from \citet{kur95}. 
 
The derived values of $T_{\rm eff}$, $\log g$ and $v\sin i$ are reported in 
Table~\ref{tab2} (columns 4,9, and 3, respectively). The table also shows the
equivalent spectral types and luminosity classes derived by comparing these values
of $T_{\rm eff}$ and $\log g$ with the tables in \citet{schmidt}. In Fig.\,\ref{spectra}, 
we show the spectra in three different wavelength ranges for two stars observed 
with both telescopes. The following lines are plotted: Mg\,\textsc{ii} $\lambda$4481\,\AA, 
Mg\,\textsc{i} $\lambda\lambda$5167\,\AA, 5172\,\AA\ and the 5183-\AA\ triplet, 
and H$\beta$. The Mg\,\textsc{i} triplet was also used to check the derived values of 
$\log g$ for the coolest stars. 

 \begin{figure} 
\includegraphics[width=8cm]{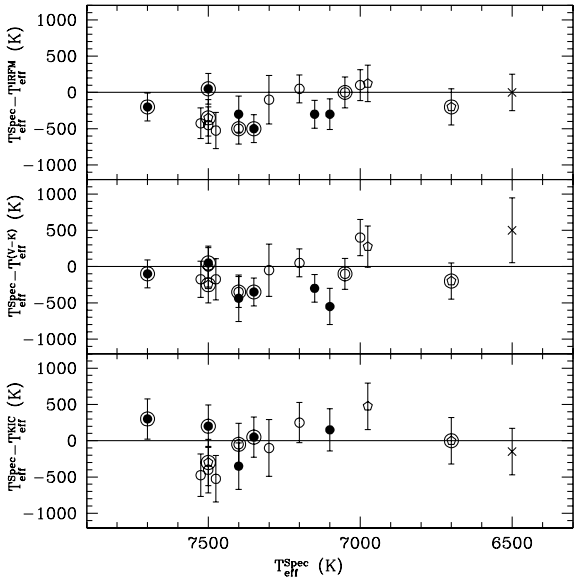} 
\caption{Comparison of $T_{\rm eff}$ obtained spectroscopically and 
photometrically via IRFM (top panel), $(V-K)$ (middle panel) and KIC
(bottom panel). 
Filled circles, pentagons and open circles represent variables
classified as pure $\delta$ Sct, pure $\gamma$ Dor, and hybrids,
respectively; the cross shows the candidate W\,Uma variable (see
section~\ref{kepler} and Table ~\ref{tab:kepler}). 
Symbols surrounded by circles refer to 
stars for which Str\"omgren photometry is available in the literature
(i.e. more precise reddening estimate). Note that for the sake of
clarity, the $T_{\rm  eff}^{\rm Spec}$ of the three stars in NGC 6866 
have been shifted by $\pm$\,25\,K. Note also that the star  
KIC\,08583770 (HD\,189177) is not visible in the figure 
because it lies outside the boundaries of the plots.} 
\label{confrTeff} 
\end{figure} 

 \begin{figure} 
\includegraphics[width=8cm]{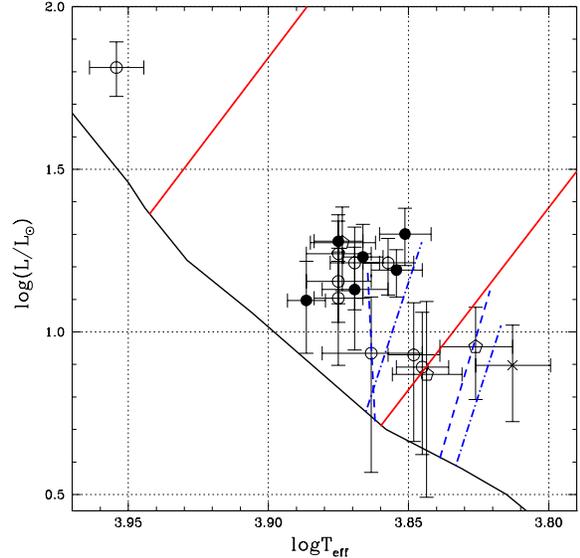} 
\caption{HR diagram for the nineteen stars investigated in this paper.
  Symbols are as in Fig.~\ref{confrTeff}. Note that the 
$T_{\rm eff}$ of KIC\,05965837 (HAT199-00623) was artificially lowered by 
25 K to avoid a complete overlap with the star KIC\,04570326 (HAT\,199-01905). 
The black solid line is the ZAMS from \citet{pickles}; the red solid lines 
show the $\delta$\,Sct instability strip by \citet{breger98}; the blue dashed 
and dotted-dashed lines show the empirical and theoretical red edge of the $\gamma$\,Dor 
instability strip by \citet{handler} and \citet{guzik}, respectively.} 
\label{hrdiagr} 
\end{figure} 
 
\subsection{Comparison between astrophysical parameters derived by different methods}
\label{individuals}

For the 15 stars with spectral types available in the literature (see column 6 of 
Table~\ref{tab1})\footnote{KIC\,05296877 (HAT\,199-27597)
and the three objects in NGC 6866 have no spectral type known 
before.} we can compare these values with those derived in the present paper 
(column 2 of Table~\ref{tab2}). For seven stars there is agreement.
For the three stars classified as metallic-lined (Am stars) by \citet{abt}, our
inferred spectral type agrees with that from Balmer lines derived by this
author on the basis of 1\,\AA\ resolution spectra. For the remaining five
stars, the discrepancy is large, with differences of more than three or four
spectral sub-types. This is not surprising because the 
nature of several classifications in the literature is uncertain or based on photometry. 
For these stars we adopt the values from our spectroscopic analysis. The only high-resolution 
study in the literature is by \citet{nordstrom} for the star KIC\,07798339 (HD\,173109). 
These authors analysed echelle spectra in the narrow wavelength range $5165.77 - 5211.25$\,\AA\ to 
obtain $T_{\rm eff}$\,=\,7000\,K, $\log g$\,=\,3.5 and $v\sin i$\,=\,15.4\,km\,s$^{-1}$ 
(no errors available). The difference of 300\,K in $T_{\rm eff}$ is not 
significant within the errors, and the $\log g$ and $v \sin i$ values are in good 
agreement with our results. 

It is useful to compare the values of $T_{\rm eff}$ derived
spectroscopically with those obtained via photometric methods 
(IRFM, (V~-~K) calibration, KIC). Inspection of Fig.~\ref{confrTeff}, which
illustrates such a comparison, shows a general good agreement among all these
values. Quantitatively, a weighted mean of such differences gives: 
$T_{\rm eff}^{\rm  Spec}-T_{\rm eff}^{\rm IRFM}$=$-$200$\pm$200; 
$T_{\rm eff}^{\rm Spec}-T_{\rm eff}^{\rm (V-K)}$=$-$130$\pm$200;
$T_{\rm eff}^{\rm Spec}-T_{\rm eff}^{\rm KIC}$=$-$50$\pm$300, non significant to 
the 1$\sigma$ level.  Similarly a very good agreement is found between $T_{\rm eff}^{\rm Spec}$ 
and $T_{\rm eff}$ estimated through $uvby\beta$ photometry (see Table~\ref{tab2}). 
However there are two exceptions to this trend:

\noindent 
{\it KIC\,05724440 (HD\,187234) -} This star shows a large difference between the 
spectroscopic T$_{\rm eff}$ and that estimated from {\it uvby$\beta$} photometry 
(800\,$\pm$\,300\,K). 
This difference is at a level of $\approx$\,2.9\,$\sigma$, and deserves some 
comments. Indeed, there is no a clear explanation for such a large 
difference. The Loiano and INAF\,-\,OACT spectra give exactly the same values for 
$T_{\rm eff}$ and $\log g$, suggesting that there could be a 
problem with the photometry. The $uvby$ values reported for this star by 
\citet{hauck} are the average of the measurements by \citet{olsen} and 
\citet{jordi}, while the $\beta$ value, on which the $T_{\rm eff}$ depends, was 
measured only by the latter authors. The two quoted $uvby$ values are slightly discrepant, 
but the difference is not large enough to change significantly the derived value 
of $\log g$. We also considered the possibility that a close companion is 
affecting the photometric measurements. We visually inspected both {\it POSS~II} 
and {\it 2MASS} images of KIC\,05724440 (HD\,187234), where the star appears to 
be isolated in the near infrared. In the optical it is surrounded by a few very 
close faint stars whose contribution can hardly be considered significant. By using 
the calibration by \citet{masana06}, the discrepancy is reduced to only 350\,K, reinforcing 
our suspicion that there is something wrong with the $uvby\beta$ photometry of this star. 

\noindent 
{\it KIC\,08583770 (HD\,189177) -} The $T_{\rm eff}$=9000$\pm$200 K derived
spectroscopically for this star is in good agreement with the
  value derived from IRFM, and consistent within the errors with the
  $T_{\rm eff}$ derived from $(V-K)$ color. However there is a large
  discrepancy with respect to the KIC estimate. This occurrence could
  perhaps be explained in terms of a visual binary, with a
companion star dimmer by 3 mag  at a distance of 0.9
arcsec. Even if nothing is known about the companion, due to the small
separation, it is likely that the photometric values are affected
by the secondary star flux.

\begin{table} 
\caption{For each star the luminosity estimated on the basis of \citet{schmidt} 
tables, the masses estimated from the evolutionary tracks with canonical and non-
canonical (i.e. with overshooting) physics, respectively. The errors on the
mass and luminosity correspond to the relative precision obtained using the
BaSTI database using non-rotating models. These errors are expected to be different
when rotation is considered in the modeling (more details in the text).} 
\begin{center} 
\begin{tabular}{lrcc} 
\hline 
\hline 
\noalign{\medskip} 
KIC   &Luminosity  & Mass Can.  &  Mass NonCan.       \\ 
\noalign{\medskip} 
       &$L$/L$_{\odot}$ & $M$/M$_{\odot}$ & $M$/M$_{\odot}$ \\ 
\noalign{\medskip} 
\hline 
\noalign{\smallskip} 
03219256     & 19.0$^{+2.9}_{-2.5}$ & 2.20$^{+0.20}_{-0.15}$ & 2.20$^{+0.10}_{-0.10}$   \\ 
03429637     & 20.0$^{+6.0}_{-6.0}$ & 3.6$^{+0.70}_{-0.60}$   & 3.2$^{+0.60}_{-0.50}$   \\ 
03437940     & 12.5$^{+4.0}_{-3,9}$ & 1.80$^{+0.30}_{-0.20}$ & 1.80$^{+0.20}_{-0.10}$   \\ 
04570326     &  7.8$^{+3.7}_{-3.6}$ & 1.70$^{+0.20}_{-0.20}$ & 1.65$^{+0.30}_{-0.15}$   \\ 
05296877     &  7.9$^{+2.6}_{-2.6}$ & 1.60$^{+0.40}_{-0.20}$ & 1.70$^{+0.25}_{-0.30}$   \\ 
05724440     & 17.0$^{+4.4}_{-4.3}$ & 2.20$^{+0.60}_{-0.40}$ & 2.30$^{+0.30}_{-0.30}$   \\ 
05965837     &  7.4$^{+5.0}_{-4.3}$ & 1.70$^{+0.30}_{-0.20}$ & 1.65$^{+0.35}_{-0.15}$   \\ 
05988140     & 16.3$^{+4.7}_{-4.6}$ & 2.00$^{+0.60}_{-0.40}$ & 2.20$^{+0.30}_{-0.30}$   \\ 
07119530     & 18.8$^{+5.4}_{-5.4}$ & 2.20$^{+0.60}_{-0.40}$ & 2.40$^{+0.40}_{-0.40}$   \\ 
07798339     &  9.0$^{+2.9}_{-2.8}$ & 1.80$^{+0.30}_{-0.20}$ & 1.90$^{+0.25}_{-0.30}$   \\ 
08197788     & 12.7$^{+4.9}_{-4.8}$ & 1.85$^{+0.45}_{-0.15}$ & 1.80$^{+0.40}_{-0.20}$   \\ 
08264404     & 17.4$^{+5.5}_{-5.2}$ & 2.30$^{+0.60}_{-0.45}$ & 2.20$^{+0.50}_{-0.40}$   \\
08264698     & 14.3$^{+3.8}_{-3.6}$ & 2.00$^{+0.30}_{-0.15}$ & 1.90$^{+0.30}_{-0.20}$   \\ 
08583770     & 65.0$^{+19.0}_{-19.0}$ & 4.80$^{+1.00}_{-0.80}$ & 4.40$^{+0.80}_{-0.60}$   \\ 
09655114     & 13.5$^{+4.7}_{-4.7}$ & 1.90$^{+0.30}_{-0.30}$ & 1.90$^{+0.40}_{-0.30}$   \\ 
09775454     &  8.5$^{+3.8}_{-3.8}$ & 1.70$^{+0.20}_{-0.20}$ & 1.65$^{+0.30}_{-0.15}$   \\ 
11402951     & 15.5$^{+2.4}_{-2.7}$ & 2.30$^{+0.20}_{-0.30}$ & 2.20$^{+0.10}_{-0.10}$   \\ 
11445913     & 16.3$^{+3.2}_{-3.3}$ & 1.95$^{+0.30}_{-0.10}$ & 2.20$^{+0.10}_{-0.10}$   \\ 
11973705     &  8.6$^{+4.2}_{-4.9}$ & 1.60$^{+0.30}_{-0.10}$ & 1.60$^{+0.30}_{-0.15}$   \\  
\noalign{\smallskip} 
\hline 
\end{tabular} 
\end{center} 
\label{tab3} 
\end{table} 

\begin{figure*} 
\includegraphics[width=8cm]{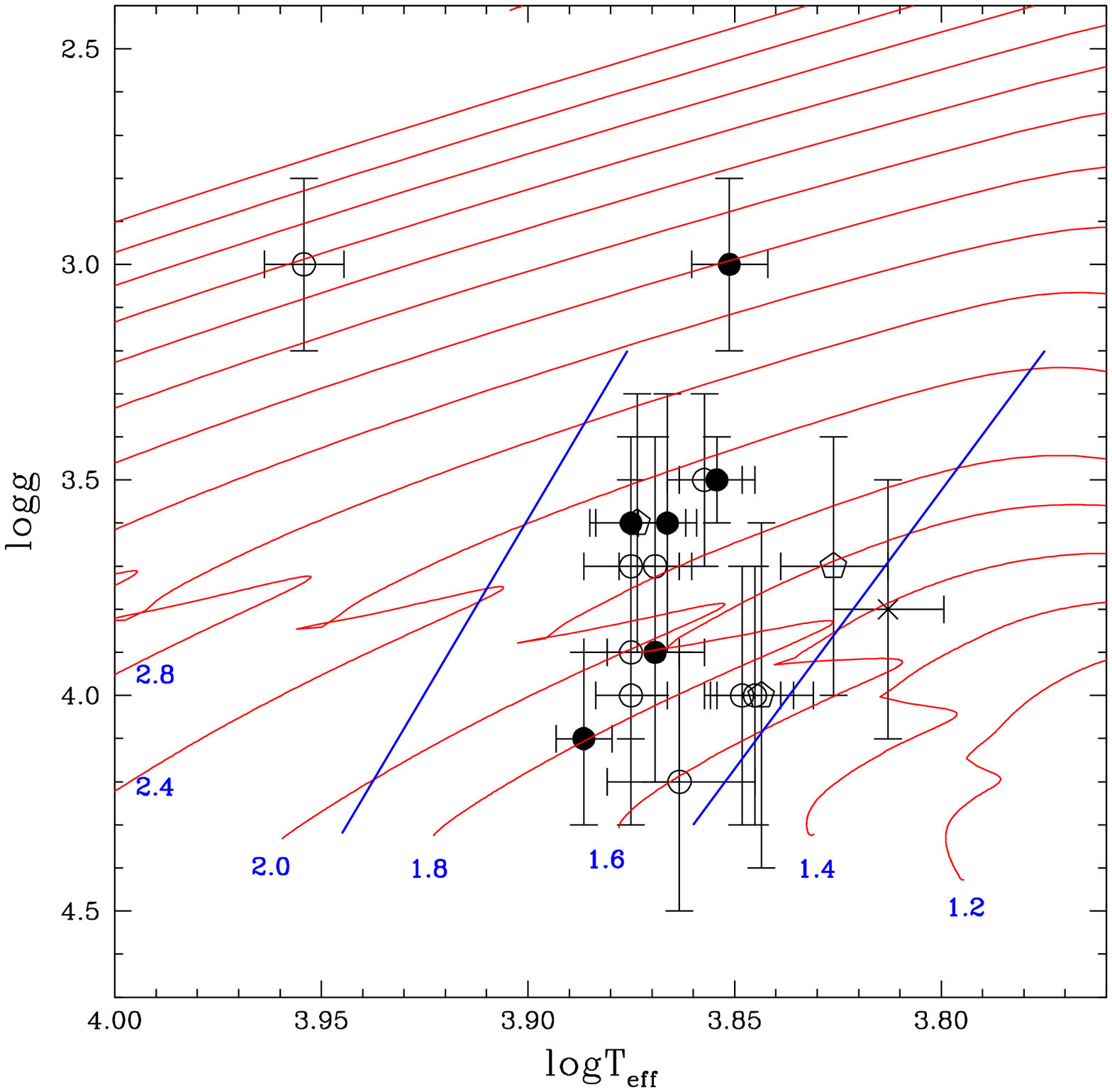} 
\includegraphics[width=8cm]{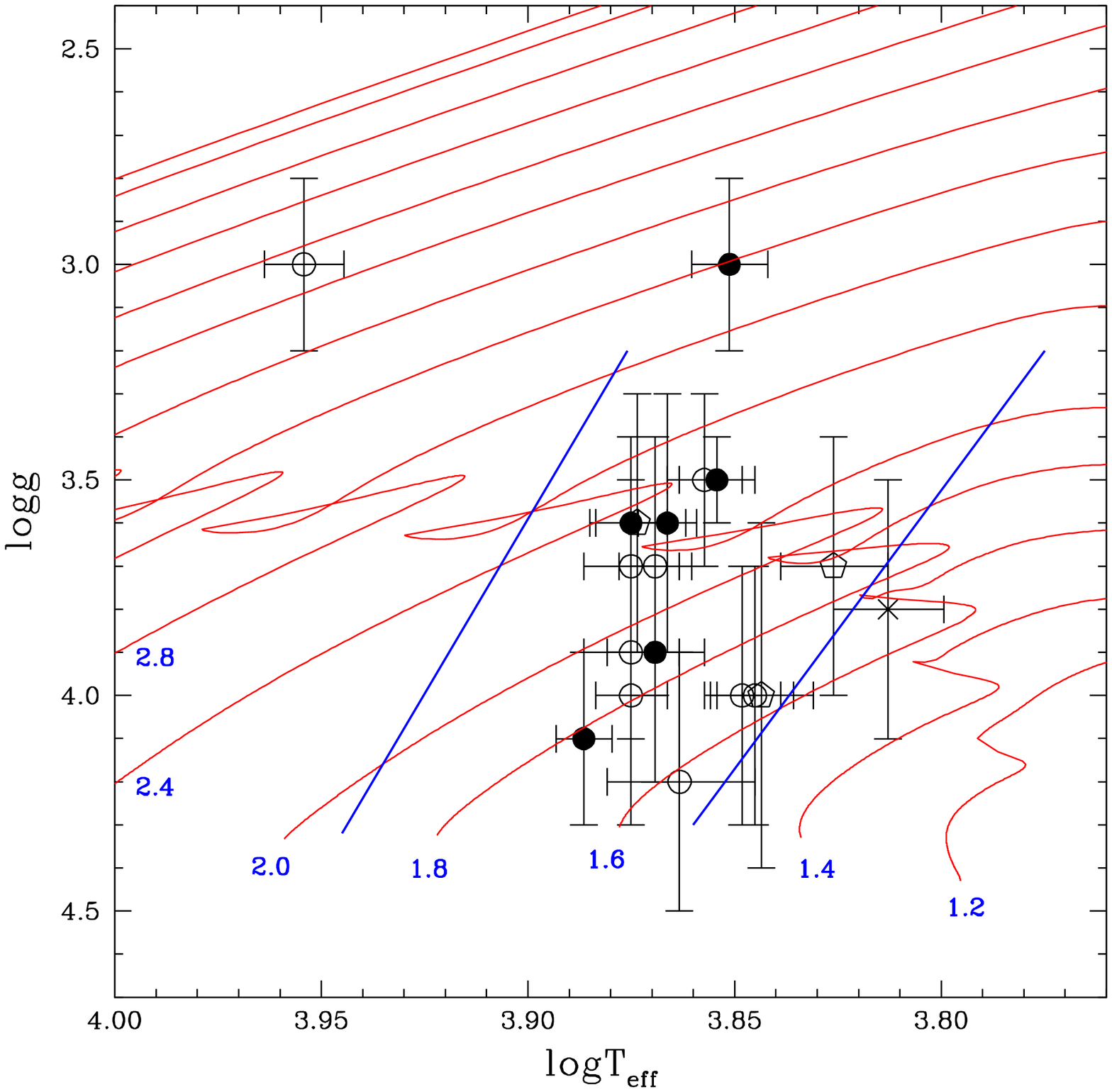} 
\caption{$\log g$-$\log T_{\rm eff}$ diagram for the investigated stars. Symbols are 
as in Fig.~\ref{confrTeff}. For comparison the instability strip for 
$\delta$\,Sct stars by \citet{breger98} is shown as straight (blue) lines. 
The evolutionary tracks for the canonical (left panel) and non-canonical (right 
panel) physics are also shown. Each track is labelled with the mass (blue number) expressed
in M$_{\odot}$.}
\label{fig3} 
\end{figure*} 
 
\section{The HR and $\log \MakeLowercase{g} - \log  T_{\rm eff}$ diagrams} 
\label{hrd} 
 
The stellar parameters $\log g$ and $\log T_{\rm eff}$
determined in the previous section 
allow us to estimate the luminosity of the investigated objects by interpolating 
the tables by \citet{schmidt}. The result is reported in Table~\ref{tab3}. Note 
that the errors on the luminosity were evaluated through the same tables by taking 
into account the errors on  $\log g$ and $\log T_{\rm eff}$. Figure\,\ref{hrdiagr} 
shows the HR diagram for the nineteen stars studied in this work, in comparison 
with the zero-age main sequence (ZAMS) \citep{pickles}, the observed instability 
strip for $\delta$\,Sct stars \citep{breger98} as well as the empirical and 
theoretical red edge of the $\gamma$\,Dor instability strip \citep{handler, guzik}. 
We note that the pulsating variables (see Section~\ref{kepler}) are in the expected
position, i.e. inside the instability strip, except for KIC\,08583770
(HD\,189177) and KIC\,05296877 (HAT\,199-27597) which are hotter and
cooler than the instability strip, respectively. The former was
discussed in the previous section, the latter is not a pulsating star
(see Section~\ref{kepler}). 
 
We can use the values of $\log g$ and $\log T_{\rm eff}$ to estimate the mass of 
the target stars. For this purpose we used the evolutionary tracks in the {\it 
BaSTI} database\footnote{http://albione.oa-teramo.inaf.it/} which are based on the 
\textsc{franec} evolutionary code \citep{chieffi}. We retrieved tracks with both 
canonical and non-canonical (i.e. with convective overshooting: $\lambda_{OV}=0.2H_p$) 
physics in the mass range $1 - 6$\,M$_{\odot}$ with $[M/H]= 0.058$, $Z= 0.0198$, 
$Y= 0.273$, and mixing lenght\,=\,1.913 \citep{pietri04}. 
In Fig.~\ref{fig3} we show the $\log g - \log T_{\rm eff}$ diagram. The left and 
right panels in the figure show canonical and non-canonical tracks, respectively. 
The resulting masses for the two cases (listed in Table\,\ref{tab3}) are very 
similar and agree well within 1$\sigma$. 
On other hand, $\delta$\,Sct stars are typically fast-rotating
objects, and the 
rotation effects on the structure and 
evolution might modify the estimates of global parameters of the stars
\citep[see e.g. ][]{goupil05, suarez05, fox06}.  To verify if this
effect is important in our case, we considered a typical case for a
star with mass 1.7-1.8 M\sun. According to \citet{suarez05}, even in 
case of v$\sin$i$\sim$150-200 km/s, the effect on the mass estimate
from the HR diagram is of few \%, well within the uncertainty due to
the errors on the empirical estimates of 
luminosity and effective temperature (see table~\ref{tab3}).

\begin{table} 
\caption{Comparison between the luminosity derived from present spectroscopy 
\changea and \citet{schmidt} tables and the one 
obtained from the HIPPARCOS parallaxes or Cluster distance (see text).} 
 \begin{center} 
 \begin{tabular}{lccr} 
 \hline 
 \hline 
 \noalign{\medskip}  
KIC  & $\pi$(mas) &  $L$/L$_{\odot}$ & $L$/L$_{\odot}$  \\  
\noalign{\medskip} 
         &                  &  Hipp. &  Spect.\\ 
\noalign{\medskip} 
\hline 
\noalign{\smallskip}  
03429637 (HD\,178875)      & 3.75$\pm$0.58 &  56.2$^{+26.3}_{-18.0}$ &20.0$^{+6.0}_{-6.0}$ \\ 
05724440 (HD\,187234)      & 8.02$\pm$0.51 &  11.1$^{+2.1}_{-1.8}$ &17.0$^{+4.3}_{-4.3}$ \\  
07798339 (HD\,173109)      & 6.86$\pm$0.48 &  13.4$^{+2.8}_{-2.3}$ & 9.0$^{+2.9}_{-2.8}$ \\  
11402951 (HD\,183489)      & 5.91$\pm$0.63 &  14.9$^{+4.4}_{-3.4}$ &15.5$^{+2.4}_{-2.7}$\\ 
\noalign{\smallskip} 
\hline 
\noalign{\smallskip}
Star  & D (pc) & $L$/L$_{\odot}$  & $L$/L$_{\odot}$  \\  
\noalign{\medskip} 
         & &Cluster &  Spect.\\ 
\noalign{\medskip} 
\hline 
\noalign{\smallskip}
08197788  (NGC6866-V1)      & 1200$\pm$120 & 10.3$^{+3.3}_{-2.5}$   &  12.7$^{+5.0}_{-4.8}$    \\ 
08264404  (NGC6866-V3)      & 1200$\pm$120 & 20.6$^{+6.5}_{-5.0}$ & 17.4$^{+5.5}_{-5.2}$  \\  
08264698  (NGC6866-V2)      & 1200$\pm$120 & 17.9$^{+5.7}_{-4.4}$  & 14.3$^{+3.8}_{-3.6}$     \\  
09655114  (NGC6811-RH35)    & 1030$\pm$50 & 17.2 $^{+5.9}_{-4.4}$  & 13.5$^{+4.7}_{-4.7}$     \\ 
\noalign{\smallskip}   
 \hline 
\end{tabular} 
\end{center} 
\label{tab4} 
\end{table}

\section{Checks of the results by means of Parallaxes and Cluster stars} 

We used the parallaxes measured by the HIPPARCOS satellite \citep{perryman97} to
verify the luminosities derived in the present work. We also estimated independently 
the luminosity of cluster stars by adopting the distances found in the literature obtained through 
e.g. isochrone fitting.
 
Only four stars in our sample are sufficiently bright for inclusion in the 
HIPPARCOS parallax catalogue. These are listed in Table\,\ref{tab4} together with 
the parallaxes from the \citet{leeuwen} revised catalogue. To derive the 
luminosity we used the $V$ and $E(B-V)$ values listed in Table\,\ref{tab1} as well 
as the bolometric correction as a function of spectral type from \citet{pickles}. 
The resulting luminosities and errors are listed in Table\,\ref{tab4} where our 
spectroscopic results are also shown for comparison purposes.  An inspection of 
the table reveals that there is agreement within the errors. The only obvious discrepancy 
is found for the star KIC\,03429637 (HD\,178875). The difference in luminosity is 
$> 1\,\sigma$, and deserves some discussion. We did not find any significant difference 
between the spectroscopic and the photometric estimates of $T_{\rm eff}$ for this star. 
Furthermore, the HIPPARCOS parallax \citep{leeuwen} is very small relative to the parallax 
estimated from its spectral type and apparent magnitude. In our opinion, KIC\,03429637 (HD\,178875) 
is very likely a double star \citep{dommanget94}. The binary nature can significantly affect 
the estimated parallax, colour, and $T_{\rm eff}$. As mentioned in Section~\ref{parameter}, 
our spectroscopic determination of $T_{\rm eff}$ is in agreement with that derived by \citet{abt}.

As for cluster stars, we have to estimate the distances to the
host clusters NGC 6866 and NGC 6811 first.  
\vskip 1mm
{\sl NGC 6866:} as reported by \citet{molenda}, both the distance
modulus and E(B-V) vary significantly from author to author. Here we
decided to assume a distance D=1200$\pm$120 pc as in  \citet{molenda} (no
error on distance is available in the literature, we assumed
conservatively an uncertainty of 10\%). As for the reddening we
adopted $E(B-V)=0.12\pm0.02$, according to \citet{dutra} who 
made a study of the foreground and background dust in the direction of the
cluster.  The resulting luminosities for the three variables in NGC
6866 are shown in Table~\ref{tab4} in comparison with our 
estimates. The agreement is good within the errors. 
\vskip 1mm
{\sl NGC 6811:} distance modulus and $E(B-V)$ of this cluster were measured by 
\citet{Glushkova} and \citet{luo}. They found $DM=10.42\pm0.03$,
$E(B-V)$=0.12$\pm$0.02,  and $DM=10.59\pm0.09$,
$E(B-V)$=0.12$\pm$0.05, respectively. To estimate the distance, we
made a weighted mean of these results, obtaining D=1030$\pm$50 pc.
Then, we calculated the luminosity for the star KIC\,09655114
(NGC6811-RH35) which is reported in the last row of
Table~\ref{tab4}. Again, we note the good agreement within the errors 
with the spectroscopic result.

\begin{figure*} 
\centering 
\includegraphics[scale=0.85]{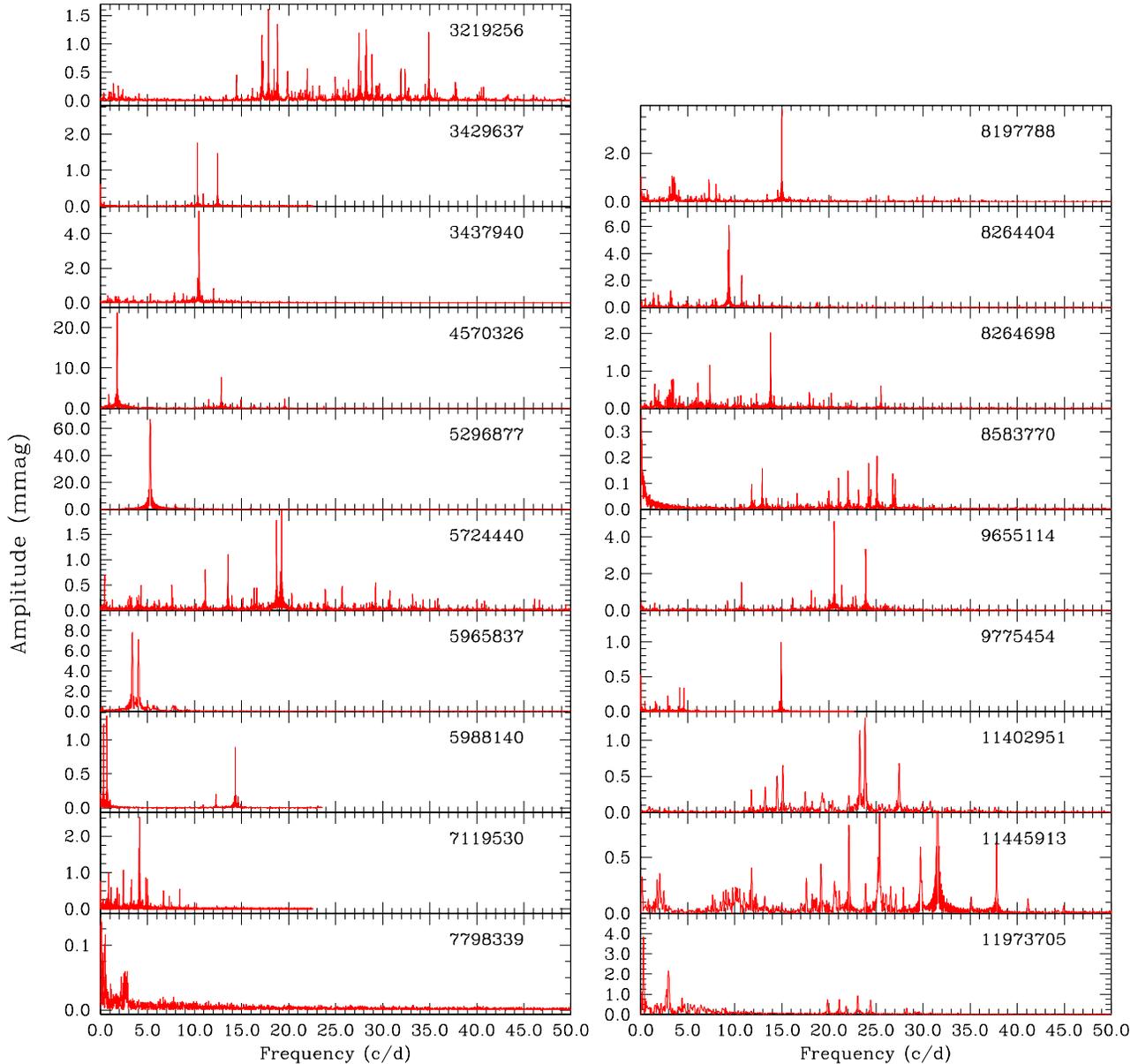} 
\caption{Periodograms of stars of Table~\ref{tab1}. The precision in amplitude of the peaks in these 
periodograms is typically in the range of $1 - 10$\,$\mu$mag.} 
\label{fig:per1} 
\end{figure*} 
 

\section{{\it Kepler} observations} 
\label{kepler} 
 
Information on {\it Kepler} observations for the stars studied here is given in 
Table\,\ref{tab:kepler}. For 15 out of 19 stars short cadence observations are 
available. From these data we calculated the periodograms (by using
L. Balona's custom software, based on a combination of FFT and normal
periodogram)  
shown in Fig.~\ref{fig:per1}. We note that practically all stars show 
peaks in both the low-frequency ($\gamma$\,Dor; g\,mode) and high-frequency 
($\delta$\,Sct; p\,mode) regions. In this sense, practically all $\delta$\,Sct 
stars observed by {\it Kepler} are hybrids. This is a surprising finding which has 
been discussed in \citet{griga10}. To make a distinction, we followed the classification 
scheme proposed by latter author. We visually classified the stars as $\delta$\,Sct if most 
of the peaks are in the $\delta$\,Sct region and as $\delta$\,Sct -- $\gamma$\,Dor if most of 
the peaks are in the $\delta$\,Sct region but with a significant contribution from the 
$\gamma$\,Dor region. The frequency 5\,c/d was taken as the boundary between  the 
two regions. Following similar arguments, we classify a star as $\gamma$\,Dor or 
$\gamma$\,Dor -- $\delta$\,Sct. There appears to be physical significance to such a scheme, as 
discussed by \citet{griga10}. We applied these classification criteria to the stars of this study, 
and classified six targets as $\delta$\,Sct stars, five stars as $\delta$\,Sct-$\gamma$\,Dor hybrids, 
four stars as $\gamma$\,Dor-$\delta$\,Sct hybrids, and three stars as pure $\gamma$\,Dor pulsators 
(see also Table 5).
Below we discuss the targets that show particularities in their
periodogram (Fig.~\ref{fig:per1}). 

{\it KIC\,03429637 (HD178875)}: As noted above, this star shows a $> 1\,\sigma$ difference between the luminosity values  
derived from the HIPPARCOS parallax and from spectroscopy (see Table\,\ref{tab4}). Binarity is a  possible  
explanation for this discrepancy. The frequency spectrum shows two dominant peaks near 10 and 12 d-1.  
A model in terms of $\delta$\,Sct pulsations, rotation and/or binarity needs to be investigated.
 
{\it KIC\,05296877 (HAT\,199-27597)}: This star is the  coolest star in the sample and lies outside the instability  
strip. The periodogram shows a single strong peak at $f = 5.302$\,c/d. KIC\,5296877 is probably a contact  
binary with an orbital period of $2/f = 0.38$\,d. The late spectral type of F4.5IV and the large value 
$v \sin i = 200$\,km\,s$^{-1}$ suggest that it is a high amplitude ellipsoidal variable.
 
{\it KIC\,07119530 (HD\,183787)}: this star has been classified as pure $\gamma$
Dor  as the frequencies are predominantly in the $\gamma$ Dor range
and  those in the  $\delta$ Sct region have  low amplitude. However
the star lies in the middle of the $\delta$ Sct instability strip. All
the temperature indicators adopted in this paper agree very well one
each other and indicate a  $T_{\rm eff}>$7500 K, i.e. a bit too hot
for a pure $\gamma$ Dor variable. We conclude that the variability classification
of this star is uncertain since it could be a $\gamma$ Dor -- $\delta$
Sct  Hybrid. 

\begin{figure} 
\centering 
\includegraphics[width=8cm]{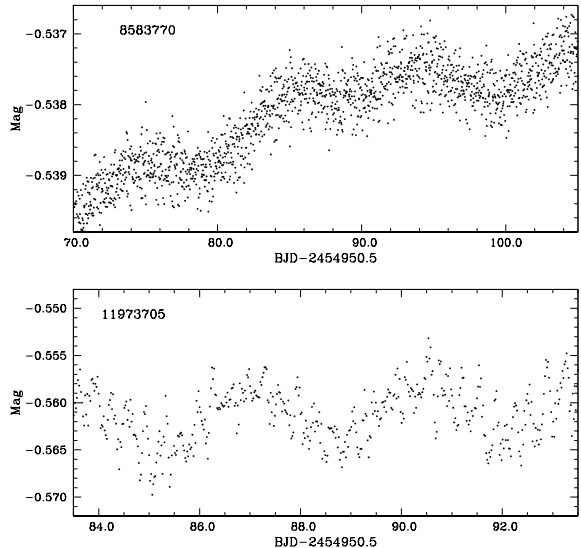} 
\caption{Top panel: light curve of K8583770 (HD\,189177). Bottom panel: light 
curve of K11973705 (HD\,234999). The time is expressed in days starting from HJD\,=\,2454950.5
and the brightness is given in mmag.} 
\label{fig:Kep} 
\end{figure}

{\it KIC\,08583770 (HD\,189177)}: This star is the hottest star in the group, and lies outside the instability 
strip (see Fig~\ref{fig3}). The periodogram shows significant power at very low frequency. The light curve 
of KIC\,08583770 is presented in Fig.\,\ref{fig:Kep}. No  specific period can be deduced, but it is clear 
that there is something  peculiar about this star, which needs further investigation.

{\it KIC\,09775454 (HD185115)}: The frequency spectrum of KIC\,09775454 shows one dominant peak in the 
$\delta$\,Sct region, and several - seemingly equidistant - peaks at lower frequencies. Further investigation  
of the {\it Kepler} light curves will clarify if  this star is a $\delta$\,Sct star with rotational  
modulation effects.
 
{\it KIC\,11973705 (HD\,234999)}: The light curves of KIC\,11973705 show a  periodic long-term behaviour, 
with $P \approx 4$\,d (Fig.~\ref{fig:Kep}). The spectral type B9, recorded in the Henry Draper catalogue, 
is a full spectral class too early compared to our classification of A9.5V. It is most likely a 
$\delta$\,Sct star in a binary system. 
 
\begin{table}
\caption{Information on {\it Kepler} photometry. The first column is the KIC
identification. In the second column the proposed classification is given. The
third column gives the number of photometric observations, $N$. In the fourth
column SC signifies short-cadence (1-min exposures) and LC long cadence
(29.4-min exposures). The last column gives the length of the data set,
$\Delta t$, in days.}
\begin{center}
\begin{tabular}{lcrrr}
\hline
\hline
\noalign{\medskip}
 KIC ID   & Type                             &        N  & Cad & $\Delta t$ (d) \\
\hline
\noalign{\smallskip}
03219256  & $\delta$\,Sct                    &   43974 & SC   &   30.03 \\
03429637  & $\delta$\,Sct                    &    9870 & LC   &  217.98 \\
03437940  & $\delta$\,Sct                    &   43989 & SC   &   30.03 \\
04570326  & $\gamma$\,Dor-- $\delta$\,Sct    &   44010 & SC   &   29.97 \\
05296877  &  -                               &   14245 & SC   &    9.72 \\
05724440  & $\delta$\,Sct                    &   43093 & SC   &   30.34 \\
05965837  & $\gamma$\,Dor                    &   14204 & SC   &    9.70 \\
05988140  & $\delta$\,Sct - $\gamma$\,Dor    &    2099 & LC   &   44.44 \\
07119530  & $\gamma$\,Dor                    &   10340 & LC   &  228.95 \\
07798339  & $\gamma$\,Dor                    &   43254 & SC   &   29.97 \\
08197788  & $\gamma$\,Dor -- $\delta$\,Sct   &   43370 & SC   &   29.97 \\
08264404  & $\delta$\,Sct -- $\gamma$\,Dor   &   43375 & SC   &   29.97 \\
08264698  & $\gamma$\,Dor -- $\delta$\,Sct   &   43348 & SC   &   29.97 \\
08583770  & $\delta$\,Sct -- $\gamma$\,Dor   &   41807 & SC   &   30.79 \\
09655114  & $\delta$\,Sct                    &   38340 & SC   &   27.11 \\
09775454  & $\delta$\,Sct -- $\gamma$\,Dor   &   10341 & LC   &  228.95 \\
11402951  & $\delta$\,Sct                    &   14240 & SC   &    9.72 \\
11445913  & $\delta$\,Sct -- $\gamma$\,Dor   &   14244 & SC   &    9.72 \\
11973705  & $\gamma$\,Dor -- $\delta$\,Sct   &   14212 & SC   &    9.72 \\
\noalign{\smallskip} \hline
\end{tabular}
\end{center}
\label{tab:kepler}
\end{table}

\section{Summary} 
\label{concl} 

We presented a spectroscopic analysis of 19 candidate $\delta$\,Sct variables 
observed by {\it Kepler} both in long and short cadence mode. The analysis is based on 
medium- to high-resolution spectra obtained at the Loiano and INAF\,-\,OACT observatories. 
For each star we derived $T_{\rm eff}$, $\log g$ and $v\sin i$ by matching the observed 
spectra with synthetic spectra computed from the {\it SYNTHE} code \citep{kur81} and using 
the LTE atmospheric models calculated by ATLAS9 \citep{kur93}. The typical errors are 
about 200\,K, 0.2\,dex, and 10 km\,s$^{-1}$ for $T_{\rm eff}$, $\log g$, and $v 
\sin i$, respectively. Equivalent spectral types and luminosity classes were also derived. 
The luminosities of the stars were obtained using the tables of \citet{schmidt}. 
 
For ten stars we used Str\"omgren photometry from the literature to estimate the 
reddening. For seven stars for which $\beta$ photometry was also available, 
$T_{\rm eff}$ and $\log g$ could be obtained for comparison with our spectroscopic 
values. In addition, $V-K$ colours, the IFRM method and values listed in the KIC were used 
to obtain independent estimates of $T_{\rm eff}$. We find a general good agreement between 
photometric and spectroscopic results.
Four stars with significant parallaxes and four cluster
member objects were used to check our estimate of the luminosities. We obtain consistent 
results for all the stars, with the exception of KIC\,03429637 (HD\,178875), which is a 
wide binary and may have an erroneus parallax determination. Moreover, for KIC\,05724440 (HD\,187234)
we suspect a problem with the $uvby\beta$ photometry, since $T_{\rm eff}$ derived from 
the $V-K$ index is in agreement with our estimate within the errors.
 
Finally, we present the periodograms for the 19 investigated stars,  based on the {\it Kepler} 
satellite photometry. These beautiful data allowed us to classify the type of variability of each 
star, including KIC\,05296877, which is a high amplitude ellipsoidal variable candidate. As a result, 
we find six pure $\delta$ Sct, 3 pure $\gamma$ Dor and nine hybrid pulsators. This classification is consistent with the derived  
physical parameters and their position in the HR diagram. As already noted by \citet{griga10}, we were 
surprised by the large number of hybrid pulsators. An asteroseismic study of these objects 
will have a strong impact on our knowledge of the evolution and internal structure of A/F stars. A more 
in depth study of the pulsational behaviour of the 18 pulsators is out of the scope of this paper, and will 
be presented in a forthcoming paper.

The stellar parameter estimates for the 18 investigated pulsating stars, presented in this work, will 
be a fundamental starting point for building proper asteroseismic models aimed at interpreting the 
frequency spectra extracted from the exceptionally good {\it Kepler} data.

\section*{Acknowledgments} 
 
This work was supported by the Italian ESS project, contract ASI/INAF
I/015/07/0, WP 03170 and by the European Helio- and Asteroseismology
Network (HELAS), a major international collaboration funded by the European 
Commission's Sixth Framework Programme. \\
Funding for the Kepler mission is provided by NASA's Science Mission
Directorate. We thank the entire Kepler team for the development and operations of this
outstanding mission.\\
This research has made use of the SIMBAD database, operated at CDS, Strasbourg, 
France. This publication makes use of data products from the Two Micron All Sky 
Survey, which is a joint project of the University of Massachusetts and the 
Infrared Processing and Analysis Center/California Institute of Technology, funded 
by the National Aeronautics and Space Administration and the National Science 
Foundation. This work has made use of BaSTI web tools. 
RSz. has been supported by the National Office for Reseach and
Technology through the Hungarian Space Office Grant No. URK09350
and the `Lend\"ulet' program of the Hungarian Academy of Sciences.
MJPFGM and AG are co-supported by project PTDC/CTE-AST/098754/2008 from FCT-Portugal.

\label{lastpage} 
 

\begin{thebibliography}{99} 
 
\bibitem[\protect\citeauthoryear{Abt}{1984}]{abt} 
 Abt, H. A., 1984, ApJ, 285, 247 
\bibitem[\protect\citeauthoryear{Blackwell \& Shallis}{1977}]{black77}
 Blackwell D.E., Shallis M.J., 1977, MNRAS 180, 177 
\bibitem[\protect\citeauthoryear{Borucki et al.}{1997}]{borucki97} 
 Borucki W.J., Koch D.G., Dunham E.W., et al., 1997, ASP Conf. Ser. 119, 153
\bibitem[\protect\citeauthoryear{Breger}{2000}]{breger00} 
 Breger M., 2000, ASP Conf. Ser. 210, 3
\bibitem[\protect\citeauthoryear{Breger \& Pamyatnykh}{1998}]{breger98} 
 Breger, M., Pamyatnykh, A. A., 1998, A\&A, 332, 958 
\bibitem[\protect\citeauthoryear{Bruntt et al.}{2008}]{bruntt08} 
 Bruntt, H., De Cat, P., Aerts, C., 2008, A\&A, 478, 487 
\bibitem[\protect\citeauthoryear{Cardelli et al.}{1989}]{cardelli89} 
 Cardelli, J. A., Clayton, G. C., Mathis, J.S., 1989, ApJ, 345, 245 
\bibitem[\protect\citeauthoryear{Chieffi \& Straniero}{1989}]{chieffi} 
 Chieffi, A., Straniero, O., 1989, ApJS, 71, 47 
\bibitem[\protect\citeauthoryear{Couteau \& Gili}{1994}]{cauteau} 
 Couteau, P., Gili, R., 1994, A\&AS, 106, 377 
\bibitem[\protect\citeauthoryear{Dommanget \& Nys}{1994}]{dommanget94} 
 Dommanget J., Nys O., 1994, Com. de l'Observ. Royal de Belgique, 115, 1 
\bibitem[\protect\citeauthoryear{Duflot et al.}{1995}]{duflot} 
 Duflot M., Figon P., Meyssonnier N. 1995, A\&AS, 114, 269 
\bibitem[\protect\citeauthoryear{Dupret et al.}{2004}]{dupret04} 
 Dupret M.-A., Grigahc\`ene A., Garrido R., Gabriel M., \& Scuflaire R., 2004, A\&A, 414, L17
\bibitem[\protect\citeauthoryear{Dutra \& Bica}{2000}]{dutra}
 Dutra C. M., Bica E., 2000, A\&A, 359, 347   
\bibitem[\protect\citeauthoryear{Fehrenbach \& Burnage}{1990}]{feh} 
 Fehrenbach, Ch., Burnage, R., 1990, A\&AS, 83, 91 
\bibitem[\protect\citeauthoryear{Fox Machado et al.}{2006}]{fox06}
 Fox Machado L., P\'erez Hern\'andez F., Su\'arez J. C., Michel E., Lebreton Y., 2006, A\&A, 446, 611 
\bibitem[\protect\citeauthoryear{Garc\'ia Hern\'andez et al.}{2009}]{garcia09}
 Garc\'ia Hern\'andez A., Moya A., Michel E., et al., 2009, A\&A, 506, 79
\bibitem[\protect\citeauthoryear{Gilliland et al.}{2010}]{gilliand10}  
 Gilliland R.L., Jenkins J.M., Borucki W.J., et al., 2010, ApJ 712, 160
\bibitem[\protect\citeauthoryear{Glushkova et al.}{1999}]{Glushkova}  
 Glushkova E.V., Batyrshinova V.M., Ibragimov M.A., 1999, Astron. Letters 25, 86
\bibitem[\protect\citeauthoryear{Goupil et al.}{2005}]{goupil05}
 Goupil M.-J., Dupret M. A., Samadi R., et al., 2005, JApA, 26, 249
\bibitem[\protect\citeauthoryear{Grigahc\`ene et al.}{2010}]{griga10} 
 Grigahc\`ene A., Antoci V., Balona L., et al., 2010, ApJL, 713, 192
\bibitem[\protect\citeauthoryear{Grigahc\`ene et al.}{2004}]{grig04} 
 Grigahc\`ene A., Dupret M.A., Garrido R., Gabriel M., Scuflaire R., 2004, CoAst, 145, 10
\bibitem[\protect\citeauthoryear{Guzik et al.}{2000}]{guzik00}
 Guzik J.A., Kaye A.B., Bradley P.A., Cox A.N., \& Neuforge C., 2000, ApJ, 542, L57
\bibitem[\protect\citeauthoryear{Handler}{2009}]{handler09}  
 Handler G., 2009, MNRAS, 398, 1339
\bibitem[\protect\citeauthoryear{Handler \& Shobbrook}{2002}]{handler} 
 Handler G., Shobbrook R. R., 2002, MNRAS, 333, 2, 251 
\bibitem[\protect\citeauthoryear{Hauck \& Mermillod}{1998}]{hauck} 
 Hauck B., Mermillod M., 1998, A\&AS, 129, 31 
\bibitem[\protect\citeauthoryear{Henry \& Fekel}{2005}]{henry05} 
 Henry G.W, Fekel F.C., 2005, AJ 129, 2026
\bibitem[\protect\citeauthoryear{Jordi et al.}{1996}]{jordi} 
 Jordi C., Figueras F., Torra J., Asiain R., 1996, A\&AS, 115, 401 
\bibitem[\protect\citeauthoryear{Kaye et al.}{1999}]{kaye99}
 Kaye A.B., Handler G., Krisciunas K., Poretti E., \& Zerbi F.M., 1999, PASP, 111, 840
\bibitem[\protect\citeauthoryear{King et al.}{2006}]{king06} 
 King H., Matthews J.M., Rowe J.F., et al. 2006, CoAst 148, 28
\bibitem[\protect\citeauthoryear{Kurucz \& Bell}{1995}]{kur95} 
 Kurucz R. L., Bell B., 1995, Kurucz CD-ROM No. 23. 
 Cambridge, Mass.: Smithsonian Astrophysical Observatory. 
\bibitem[\protect\citeauthoryear{Kurucz}{1993a}]{kur93} 
 Kurucz R.L., 1993, A new opacity-sampling model atmosphere 
 program for arbitrary abundances. In: Peculiar versus normal phenomena in 
 A-type and related stars, IAU Colloquium 138, M.M. Dworetsky, F. Castelli, 
 R. Faraggiana (eds.), A.S.P Conferences Series Vol. 44, p.87
\bibitem[\protect\citeauthoryear{Kurucz}{1993b}]{kur93b} 
 Kurucz R.L., 1993, Kurucz CD-ROM 13: ATLAS9, SAO, Cambridge, USA 
\bibitem[\protect\citeauthoryear{Kurucz \& Avrett }{1981}]{kur81} 
 Kurucz R.L., Avrett E.H., 1981, SAO Special Rep., 391 
\bibitem[\protect\citeauthoryear{Latham et al.}{2005}]{latham05} 
 Latham D.W., Brown T.M., Monet D.G., Everett M., Esquerdo G. A., Hergenrother C. W., 2005, AAS 37, 1340
\bibitem[\protect\citeauthoryear{Lindoff}{1972}]{lindoff} 
 Lindoff, U. 1972, A\&A, 16, 315 
\bibitem[\protect\citeauthoryear{Luo et al.}{2009}]{luo} 
 Luo Y. P., Zhang X. B., Luo C. Q., Deng L. C., Luo Z. Q., 2009, NewA, 14, 584
\bibitem[\protect\citeauthoryear{Masana et al.}{2006}]{masana06} 
 Masana E., Joedi C., Ribas I., 2006, A\&A, 450, 735 
\bibitem[\protect\citeauthoryear{Molenda-\.Zakowicz}{2009}]{molenda} 
 Molenda-\.Zakowicz J., Kopacki G., Ste\'slicki M., Narwid A., 2009, AcA, 59, 193
\bibitem[\protect\citeauthoryear{Moon}{1985}]{moon} 
 Moon, T.T., 1985, Ap\&SS, 117, 261 
\bibitem[\protect\citeauthoryear{Moon \& Dworetsky}{1985}]{moondwo} 
 Moon, T. T., Dworetsky, M. M., 1985, MNRAS, 217, 305 
\bibitem[\protect\citeauthoryear{Nordstrom et al.}{1997}]{nordstrom} 
 Nordstrom, B., Stefanik, R. P., Latham, D. W., Andersen, J., 1997, A\&AS, 126, 21 
\bibitem[\protect\citeauthoryear{Olsen}{1983}]{olsen} 
 Olsen, E.H., 1983, A\&AS, 54, 550 
\bibitem[\protect\citeauthoryear{Perryman et al.}{1997}]{perryman97}  
 Perryman, M.A.C., Lindegren, L., Kovalevsky, J., et al. 1997, A\&A 323, L49
\bibitem[\protect\citeauthoryear{Pickles}{1998}]{pickles} 
 Pickles A. J., 1998, PASP, 110, 863 
\bibitem[\protect\citeauthoryear{Pietrinferni et al.}{2004}]{pietri04} 
 Pietrinferni A., Cassisi S., Salaris M., Castelli F., 2004, ApJ, 612 168
\bibitem[\protect\citeauthoryear{Rowe et al.}{2006}]{rowe06} 
 Rowe J.F., Matthews J. M., Cameron C., et al. 2006, CoAst 148, 34
\bibitem[\protect\citeauthoryear{Scargle}{1982}]{scargle82} 
 Scargle, J. D. 1982, ApJ 263, 835
\bibitem[\protect\citeauthoryear{Schmidt-Kaler}{1982}]{schmidt} 
 Schmidt-Kaler T., 1982, in ``Landolt-B\"ornstein'', Vol 2b Group IV, Springer-Verlag 
\bibitem[\protect\citeauthoryear{Skrutskie et al.}{1996}]{2mass} 
 Skrutskie, M. F., Cutri, R.M., Stiening, R., Weinberg, M.D., Schneider, S. et al., 2006, AJ, 131, 1163 
\bibitem[\protect\citeauthoryear{Su\'arez et al.}{2005}]{suarez05}
 Su\'arez J.-C., Bruntt H., Buzasi D., 2005, A\&A, 438, 633
\bibitem[\protect\citeauthoryear{Uytterhoeven et al.}{2008a}]{uytte08} 
 Uytterhoeven K., Mathias P., Poretti E., et al., 2008, A\&A 489, 1213
\bibitem[\protect\citeauthoryear{Uytterhoeven et al.}{2010a}]{uytte10a}
 Uytterhoeven K., Briquet M., Bruntt H., et al., 2010, AN, submitted (arXiv:1003.6093)
\bibitem[\protect\citeauthoryear{Uytterhoeven et al.}{2010b}]{uytte10b}  
 Uytterhoeven, K., Szabo, R., Southworth, J., et al., 2010, AN, submitted (arXiv:1003.6089)
\bibitem[\protect\citeauthoryear{van Leeuwen}{2007}]{leeuwen} 
 van Leeuwen, F., 2007, A\&A, 474, 653 
\bibitem[\protect\citeauthoryear{Voigt}{2006}]{born06} 
 Voigt, H. H., 2006, in ``Landolt-B\"ornstein'', Vol 3b Group VI, Springer-Verlag 
\bibitem[\protect\citeauthoryear{Warner et al.}{2003}]{guzik} 
 Warner, P. B., Kaye, A. B.; Guzik, J. A., 2003, ApJ, 593, 1049 
\end{thebibliography}
\end{document}